# Radiation Pressure and Photon Momentum in Negative-Index Media

Masud Mansuripur[†] and Armis R. Zakharian[‡]

[†]College of Optical Sciences, The University of Arizona, Tucson, Arizona 85721

[‡]Corning Incorporated, Science and Technology Division, Corning, New York 14831

**Abstract**. Radiation pressure and photon momentum in negative-index media are no different than their counterparts in ordinary (positive-index) materials. This is because the parameters responsible for these properties are the admittance $\sqrt{\varepsilon/\mu}$ and the group refractive index $n_g$ of the material (both positive entities), and not the phase refractive index $n = \sqrt{\mu\varepsilon}$, which is negative in negative-index media. One approach to investigating the exchange of momentum between electromagnetic waves and material media is via the Doppler shift phenomenon. In this paper we use the Doppler shift to arrive at an expression for the radiation pressure on a mirror submerged in a negative-index medium. In preparation for the analysis, we investigate the phenomenon of Doppler shift in various settings, and show the conditions under which a so-called "inverse" Doppler shift could occur. We also argue that a recent observation of the inverse Doppler shift upon reflection from a negative-index medium cannot be correct, because it violates the conservation laws.

**1. Introduction**. A negative-index medium (NIM) is a natural or artificial material whose dispersion relation, $\omega(k)$, exhibits a negative slope, $d\omega/dk$, over a certain frequency interval. Here $\omega$ is the angular frequency and $k$ the wave-number of a plane electromagnetic wave propagating within the medium. Since the phase refractive index is given by $n(\omega) = ck(\omega)/\omega$, whereas the group index is defined as $n_g(\omega) = c/[d\omega(k)/dk]$, if we follow tradition and choose a positive sign for the group index, then the phase index will turn out to be negative. The negative-index could come about as a result of the nano-structure of a meta-material (e.g., photonic crystal band structure), or it could be due to the material's permittivity $\varepsilon(\omega)$ and permeability $\mu(\omega)$ both being real-valued and negative over some frequency range. In the latter case, the phase refractive index $n = \sqrt{\mu\varepsilon}$ will be negative, whereas $n_g$ as well as the admittance $\eta = \sqrt{\varepsilon/\mu}$ and impedance $Z = \sqrt{\mu/\varepsilon}$ of the medium will turn out to be positive.

It has been suggested that the Doppler shift associated with negative-index media could be the reverse of that observed in ordinary materials [1]. Also, anomalous behavior for radiation pressure and photon momentum inside negative-index media has been predicted [2]. In this paper we examine the Doppler shift from multiple perspectives, using special relativity as well as the conservation laws of energy and momentum to derive expressions for the Doppler shift under various circumstances. We show that negative-index media, by and large, behave similarly to their positive-index counterparts, and that a recent claim of experimental observation of the "inverse" Doppler shift upon reflection from a NIM is patently absurd [3]. An exception to the above rule occurs in the case of reflection from a mirror submerged in a negative-index dielectric, where, as Veselago originally suggested [1], an inverse Doppler shift is indeed plausible.

Photon momentum and radiation pressure are intimately tied to the magnitude and sign of the Doppler shift. We examine the connection between these two phenomena in some detail, and derive expressions for the radiation pressure that are fully consistent with those obtained using the generalized Lorentz force in conjunction with Maxwell's macroscopic equations [4,5]. It will be shown that, as far as radiation pressure is concerned, the relevant material parameters are its admittance $\eta(\omega)=\sqrt{\varepsilon/\mu}$ and its group refractive index $n_g(\omega)$, both of which, unlike the phase refractive index $n(\omega)$, are positive for NIMs.

**2. Doppler shift upon reflection from a mirror moving in vacuum**. Let a perfect reflector have mass $M_o$ and initial velocity $V_i$ along the $z$-axis, as shown in Fig. 1. A photon of frequency $f$, also propagating along the $z$-axis, is normally-incident on the mirror. Upon reflection, the mirror will have acquired the velocity $V_f$ and the photon frequency shifted to $f'$. We may use conservation of energy and momentum to derive the formula for the Doppler shift $\Delta f = f - f'$. In the following we will derive expressions for $f'/f$ using first the non-relativistic approach and then the relativistic approach.

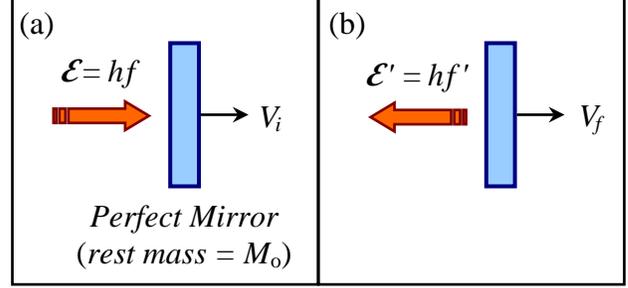

**Fig. 1**. A photon of energy $hf$ is normally incident on a perfect reflector of rest mass $M_o$ and initial velocity $V_i$. After reflection, the photon's energy is $hf'$ and the mirror's velocity is $V_f$. Conservation of energy and momentum may be used to determine the Doppler-shifted frequency $f'$ and the final velocity $V_f$ in terms of $f$, $V_i$ and $M_o$.

*Case I. Non-relativistic calculation*: Considering that the photon energy and momentum in vacuum are given by $hf$ and $hf/c$, respectively, where $h$ is Planck's constant and $c$ the speed of light in free space, we have

$$\text{Conservation of energy:} \quad h(f-f') = \tfrac{1}{2}M_o(V_f^2 - V_i^2), \tag{1a}$$

$$\text{Conservation of momentum:} \quad h(f+f')/c = M_o(V_f - V_i). \tag{1b}$$

Dividing the first of the above equations by the second yields

$$\frac{f-f'}{f+f'} = \frac{V_f + V_i}{2c} \quad \rightarrow \quad \frac{f'}{f} = \frac{1-(V_f+V_i)/2c}{1+(V_f+V_i)/2c} \approx 1 - (V_f+V_i)/c. \tag{2}$$

*Case II. Relativistic calculation*: Using the standard definition $\gamma = 1/\sqrt{1-(V/c)^2}$, we write

$$\text{Conservation of energy:} \quad h(f-f') = (\gamma_f - \gamma_i)M_o c^2, \tag{3a}$$

$$\text{Conservation of momentum:} \quad h(f+f')/c = M_o(\gamma_f V_f - \gamma_i V_i). \tag{3b}$$

Dividing the first of the above equations by the second now yields

$$\frac{f-f'}{f+f'} = \frac{(\gamma_f - \gamma_i)c}{(\gamma_f V_f - \gamma_i V_i)} \quad \rightarrow \quad \frac{f'}{f} = \sqrt{\frac{(1-V_i/c)(1-V_f/c)}{(1+V_i/c)(1+V_f/c)}}. \tag{4}$$

The results of Case I and Case II are, of course, essentially the same for a massive mirror. Also, when $M_o \to \infty$, $V_f \to V_i$ and $\Delta f = f - f' \to 2(V_i/c)f$. The reflected light will be red-shifted for a mirror moving away from the source, and blue-shifted for a mirror moving toward the source. There will be no "inverse" Doppler shifts irrespective of the nature of the reflector; in particular, the sign of the Doppler shift will not depend on whether or not the reflector is a negative-index medium.

We now use an entirely different method to obtain an expression for the Doppler shift in the system of Fig.1. This is an approximate method which works well at non-relativistic velocities. Suppose the incident plane-wave has an *E*-field amplitude given by

$$\boldsymbol{E}^{(i)}(z,t) = \boldsymbol{E}_0 \exp[\mathrm{i}(\boldsymbol{k}\cdot\boldsymbol{r}-\omega t)] = \boldsymbol{E}_0 \exp[\mathrm{i}(\omega/c)(z-ct)]. \tag{5}$$

Here $\omega = 2\pi f$ is the angular frequency of the incident beam, and $\boldsymbol{k} = (\omega/c)\hat{\boldsymbol{z}}$ is its wave-vector. If the mirror happens to be stationary and located at $z = z_o$, the reflected *E*-field amplitude will be

$$\boldsymbol{E}^{(r)}(z,t) = \boldsymbol{E}_0^{(r)} \exp\{\mathrm{i}(\omega/c)[(2z_o - z) - ct]\}. \tag{6}$$

Suppose now that the mirror is moving at constant velocity *V* along the *z*-axis. Setting $z_o = Vt$ in Eq.(6) and combining the terms that contain *t*, we find

$$\boldsymbol{E}^{(r)}(z,t) = \boldsymbol{E}_0^{(r)} \exp\{-\mathrm{i}[(\omega/c)z + (1-2V/c)\omega t]\}. \tag{7}$$

The frequency of the reflected beam is clearly given by $\omega' = (1-2V/c)\omega$, as before. Returning now to the conservation of energy expression in Eq.(1a), if we set the initial velocity $V_i$ of the mirror to zero, and assume that the effective Doppler shift is caused by the average mirror velocity ½$V_f$, the reduction in the photon energy upon reflection from an initially stationary mirror will, according to Eq.(7), be $\hbar(\omega - \omega') = (V_f/c)\hbar\omega$. This must be equal to the increase in the mirror's kinetic energy ½$M_o V_f^2$. Consequently, $M_o V_f = 2\hbar\omega/c$, that is, the momentum acquired by a massive, stationary mirror will be $2\hbar\omega/c$, which must be the difference between photon momenta before and after reflection. In this way, we have ascertained that the photon momentum in vacuum is indeed given by $\hbar\omega/c$. In what follows, we will use similar arguments to determine the radiation pressure on a mirror submerged in a transparent medium.

**3. Doppler shift at oblique incidence**. Figure 2 shows a massive mirror with a fixed orientation moving at constant velocity *V* along the *z*-axis. A photon of energy $\hbar\omega_0$ whose *k*-vector, confined to the *xz*-plane, makes an angle $\theta_0$ with the *z*-axis is reflected from the mirror and continues to propagate in the *xz*-plane, albeit at a different angle $\theta_1$ from the *z*-axis. The photon energy after reflection is $\hbar\omega_1$. The mirror is stationary in the *x'y'z'* frame, but in the *xyz* frame it moves with constant velocity *V* along the *z*-axis.

The exponential phase-factor associated with a plane-wave of frequency $\omega$ propagating in free space along its *k*-vector is $\exp[\mathrm{i}(\boldsymbol{k}\cdot\boldsymbol{r}-\omega t)]$, where $\boldsymbol{k}=(\omega/c)\hat{\boldsymbol{k}}$, with $\hat{\boldsymbol{k}}$ being a unit-vector along the propagation direction. For the incident wave, we have $\boldsymbol{k}\cdot\boldsymbol{r}-\omega t = (\omega_0/c)(x\sin\theta_0 + z\cos\theta_0 - ct)$, which in the *x'y'z'* frame becomes $(\omega_0/c)[x'\sin\theta_0 + \gamma(z'+Vt')\cos\theta_0 - c\gamma(t'+Vz'/c^2)]$. The coefficient of *t'* is thus found to be $\omega' = \gamma[1-(V/c)\cos\theta_0]\omega_0$, and this is the frequency of the incident wave as seen by the stationary mirror in its own rest frame. The reflected beam will have essentially the same frequency $\omega'$ in the rest frame of the mirror (provided $M_o$ is large), but its direction will be given by the angle $\theta'_1$. The phase factor of the reflected beam will thus be

$$\boldsymbol{k}\cdot\boldsymbol{r} - \omega t = (\omega'/c)(x'\sin\theta'_1 + z'\cos\theta'_1 - ct'),$$

which, in the *xyz* frame, may be written as

$$(\omega'/c)[x\sin\theta'_1 + \gamma(z-Vt)\cos\theta'_1 - c\gamma(t-Vz/c^2)].$$

We find, after straightforward algebraic manipulations, that

$$\tan\theta_1 = \frac{\sin\theta_1'}{\gamma[\cos\theta_1' + (V/c)]}, \tag{8a}$$

$$\omega_1 = \omega_0[1-(V/c)\cos\theta_0][1+(V/c)\cos\theta_1']/[1-(V/c)^2]. \tag{8b}$$

For non-relativistic mirror velocities (i.e., $V \ll c$), $\theta_1' \approx \theta_1$, and $\omega_1 - \omega_0 \approx (V/c)(\cos\theta_1 - \cos\theta_0)\omega_0$. The Doppler shift is clearly independent of the nature of the reflector, depending only on the velocity $V$ of the mirror and the angles $\theta_0$ and $\theta_1$ that the incident and reflected beams make with the $z$-axis. In particular, whether or not the reflector is a negative-index material is irrelevant. If the mirror is moving away from the source, the reflected light will be red-shifted, and if the mirror is moving toward the source, the reflected light will be blue-shifted, in accordance with the above expression for the Doppler shift. The Shanghai group's experiments [3], which "reveal" an inverse Doppler shift under conditions depicted in Fig. 2 are therefore meaningless. The reason these researchers found an inverse Doppler shift in their experiments can be traced to the fact that they were comparing the light reflected from a NIM with that reflected from a reference mirror. Both mirrors were moving at the same velocity and in the same direction. However, while the two reflected beams travelled in the same direction (i.e., $\theta_1$ for test beam was the same as that for reference beam), the incidence angles $\theta_0$ for the two beams were different. By adjusting the orientation of the reference mirror and, thereby, adjusting $\theta_0$ for the reference beam, one could get either sign (positive or negative) for the Doppler shift in the Shanghai experiments.

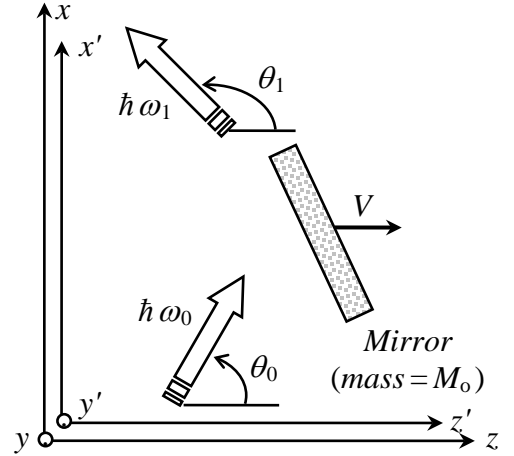

**Fig. 2**. Reflection of a photon of frequency $\omega_0$ from a moving mirror. The mirror has a large mass $M_o$ and moves with velocity $V$ along the $z$-axis. The reflected photon is Doppler-shifted to frequency $\omega_1$. In the $xyz$ frame, the propagation directions are denoted by $\theta_0$ and $\theta_1$. In the rest frame $x'y'z'$ of the mirror, where both photons have the same frequency $\omega'$, their propagation directions (not shown) are $\theta_0'$ and $\theta_1'$.

An inverse Doppler shift in free space would, in fact, contradict the laws of conservation of energy and momentum. To see this, observe that in the $xyz$ frame of Fig. 2 we may assume that the mirror gains a very small velocity $V_x\hat{x} + V_z\hat{z}$ upon the reflection of an incident photon. Working in the non-relativistic regime and assuming a very large mass $M_o$ for the mirror, we write

Momentum conservation along $x$: $(\hbar\omega_0/c)\sin\theta_0 = (\hbar\omega_1/c)\sin\theta_1 + M_o V_x,$ (9a)

Momentum conservation along $z$: $(\hbar\omega_0/c)\cos\theta_0 + M_o V = (\hbar\omega_1/c)\cos\theta_1 + M_o(V+V_z),$ (9b)

Energy conservation: $\hbar\omega_0 + \tfrac{1}{2}M_o V^2 = \hbar\omega_1 + \tfrac{1}{2}M_o[(V+V_z)^2 + V_x^2].$ (9c)

Substitution for $V_x$ and $V_z$ from Eqs. (9a) and (9b) into Eq. (9c) yields

$$\omega_1^2 - 2\{\omega_0\cos(\theta_0-\theta_1) - (M_o c^2/\hbar)[1-(V/c)\cos\theta_1]\}\omega_1 + \{\omega_0^2 - (2M_o c^2\omega_0/\hbar)[1-(V/c)\cos\theta_0]\} = 0. \tag{10}$$

Of the two solutions to the above quadratic equation in $\omega_1$, only one is acceptable, namely,

$$\omega_1 = \{\omega_0 \cos(\theta_0 - \theta_1) - (M_o c^2/\hbar)[1-(V/c)\cos\theta_1]\} + (M_o c^2/\hbar)[1-(V/c)\cos\theta_1]$$

$$\times \sqrt{1 + \frac{(2M_o c^2 \omega_0/\hbar)[1-(V/c)\cos\theta_0] - 2\omega_0 \cos(\theta_0-\theta_1)(M_o c^2/\hbar)[1-(V/c)\cos\theta_1] - \omega_0^2 \sin^2(\theta_0-\theta_1)}{(M_o c^2/\hbar)^2 [1-(V/c)\cos\theta_1]^2}}$$

$$\approx \{\omega_0 \cos(\theta_0 - \theta_1) - (M_o c^2/\hbar)[1-(V/c)\cos\theta_1]\} + (M_o c^2/\hbar)[1-(V/c)\cos\theta_1]$$

$$\times \left\{1 + \frac{(2M_o c^2 \omega_0/\hbar)[1-(V/c)\cos\theta_0] - 2\omega_0 \cos(\theta_0-\theta_1)(M_o c^2/\hbar)[1-(V/c)\cos\theta_1] - \omega_0^2 \sin^2(\theta_0-\theta_1)}{2(M_o c^2/\hbar)^2 [1-(V/c)\cos\theta_1]^2}\right\}$$

$$\approx \cancel{\omega_0 \cos(\theta_0-\theta_1)} + \frac{1-(V/c)\cos\theta_0}{1-(V/c)\cos\theta_1}\omega_0 - \cancel{\omega_0 \cos(\theta_0-\theta_1)} - \underbrace{\frac{\omega_0^2 \sin^2(\theta_0-\theta_1)}{2(M_o c^2/\hbar)[1-(V/c)\cos\theta_1]}}_{\approx 0}$$

$$\approx \frac{1-(V/c)\cos\theta_0}{1-(V/c)\cos\theta_1}\omega_0. \tag{11}$$

Once again, we simplify the above equation in the non-relativistic limit ($V \ll c$) to arrive at the same result as before, namely,

$$\omega_1 - \omega_0 \approx (V/c)(\cos\theta_1 - \cos\theta_0)\omega_0. \tag{12}$$

Observation of an inverse Doppler shift from a NIM in the system of Fig. 2 is thus impossible, as it would violate the energy and momentum conservation laws.

**4. Doppler shift inside a transparent slab**. Let a long and wide light pulse of frequency $\omega_0$, $E$-field amplitude $E_0 \hat{x}$, and $H$-field amplitude $H_0 \hat{y}$, propagate in free space along the $z$-axis, as shown in Fig. 3. The exponent of the phase-factor $\exp[i(\mathbf{k} \cdot \mathbf{r} - \omega t)]$ is $(\omega_0/c)(z-ct)$ in the $xyz$ frame and $(\omega_0/c)[\gamma(z'+Vt') - c\gamma(t'+Vz'/c^2)] = \gamma(1-V/c)(\omega_0/c)(z'-ct')$ in the $x'y'z'$ frame. Thus, in the slab's rest frame, the receding light source appears to have a frequency $\omega' = \sqrt{(1-V/c)/(1+V/c)}\,\omega_0$. The $E$- and $H$-fields of the incident pulse may also be Lorentz-transformed into the $x'y'z'$ frame, yielding $E_0' = \gamma(1-V/c)E_0 = \sqrt{(1-V/c)/(1+V/c)}\,E_0$, and likewise, $H_0' = \sqrt{(1-V/c)/(1+V/c)}\,H_0$. It is clear that the Poynting vector $\mathbf{E}_0 \times \mathbf{H}_0$ has reduced in magnitude by a factor $(1-V/c)/(1+V/c)$, while the frequency spectrum of the pulse has contracted by the square root of this factor. The light pulse, therefore, is longer in the $x'y'z'$ frame by the same square-root factor, yielding a total pulse energy that is reduced by the Doppler factor $\sqrt{(1-V/c)/(1+V/c)}$. Considering that, in going from one frame to the other, the number of photons in the pulse remains intact, the change in the total pulse energy must be the same as the change in its frequency $\omega_0$, and this indeed is seen to be the case.

Next, we consider the Fresnel reflection coefficient at the entrance facet of the slab, which, in the slab's rest frame, is $r(\omega') = [1 - \sqrt{\varepsilon(\omega')/\mu(\omega')}]/[1 + \sqrt{\varepsilon(\omega')/\mu(\omega')}]$. The reflected $E$-field and $H$-field amplitudes will be multiplied by $r(\omega')$ and, upon transforming to the $xyz$ frame, they become

$$\mathbf{E}_0^{(r)} = r(\omega')\left[\frac{1-(V/c)}{1+(V/c)}\right]E_0 \hat{x}, \tag{13a}$$

$$\boldsymbol{H}_0^{(r)} = -r(\omega')\left[\frac{1-(V/c)}{1+(V/c)}\right]H_0\hat{\boldsymbol{y}}. \tag{13b}$$

It is also straightforward to Lorentz transform the phase-factor of the reflected beam into the $xyz$ frame, upon which we find the reflected frequency to be $\omega_r = [(1-V/c)/(1+V/c)]\omega_0$. In other words, the number of reflected photons is $r^2(\omega')$ times the number of incident photons, while the energy of each reflected photon is reduced by the Doppler factor $(1-V/c)/(1+V/c)$. Note that these results are independent of whether the transparent slab is positive- or negative-index, as the admittance $\sqrt{\varepsilon(\omega')/\mu(\omega')}$ appearing in the expression of $r(\omega')$ is positive for both types of media.

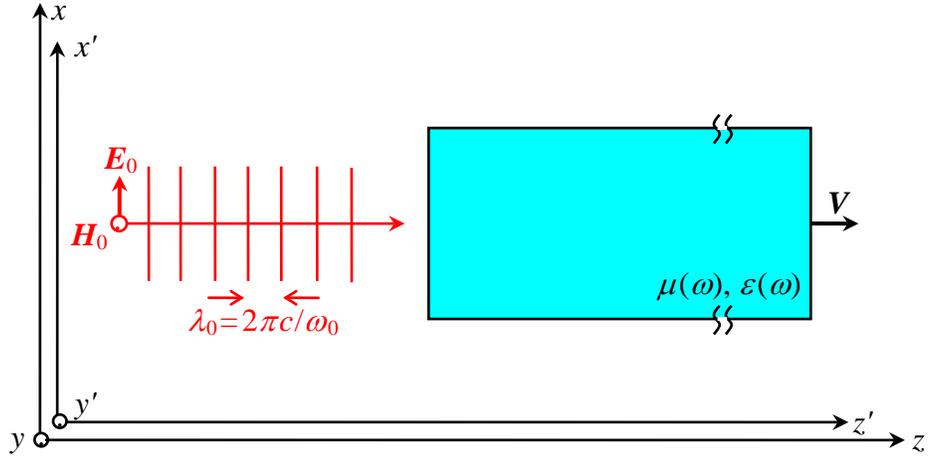

**Fig. 3**. A long and wide pulse of light having frequency $\omega_0$, $E$-field amplitude $E_0\hat{\boldsymbol{x}}$, and $H$-field amplitude $H_0\hat{\boldsymbol{y}}$, propagates in free space along the $z$-axis. The pulse enters a transparent slab of a magnetic dielectric specified by its real-valued permeability $\mu(\omega)$ and permittivity $\varepsilon(\omega)$. In the $xyz$ frame, the source of the light is stationary, while the slab moves at constant velocity $V$ along the $z$-axis. The slab is at rest in the $x'y'z'$ frame.

It remains to investigate the plane-wave that propagates inside the slab. In the slab's rest frame, this beam has frequency $\omega'$, $E$-field amplitude $E_0'^{(t)} = [1+r(\omega')]\sqrt{(1-V/c)/(1+V/c)}\,E_0$, and $H$-field amplitude $H_0'^{(t)} = [1-r(\omega')]\sqrt{(1-V/c)/(1+V/c)}\,H_0$, where the superscript ($t$) stands for "transmitted." The corresponding $D$- and $B$-fields are given by $D_0'^{(t)} = \varepsilon_0\varepsilon(\omega')E_0'^{(t)}$ and $B_0'^{(t)} = \mu_0\mu(\omega')H_0'^{(t)}$.

The exponent of the phase-factor inside the slab is given by $(\omega'/c)[n(\omega')z'-ct']$, where the phase refractive index $n(\omega') = \sqrt{\mu(\omega')\varepsilon(\omega')}$ may be positive or negative, depending on the medium being positive- or negative-index. Lorentz transformation of this phase-factor to the $xyz$ coordinates yields

$$(\omega'/c)[n(\omega')z'-ct'] \to \sqrt{(1-V/c)/(1+V/c)}\,(\omega_0/c)[n(\omega')\gamma(z-Vt)-c\gamma(t-Vz/c^2)]$$

$$= \left[\frac{1+n(\omega')(V/c)}{1+(V/c)}\right](\omega_0/c)\left[\frac{n(\omega')+(V/c)}{1+n(\omega')(V/c)}z-ct\right]. \tag{14}$$

Clearly, the frequency of the beam inside the slab is $\omega^{(t)} = \omega_0[1+n(\omega')(V/c)]/[1+(V/c)]$, and the effective refractive index of the medium, as seen by a stationary observer in the $xyz$ frame, is $[n(\omega')+(V/c)]/[1+n(\omega')(V/c)]$. At non-relativistic speeds, the Doppler shift inside the slab will be

$$\omega^{(t)} - \omega_0 \approx [n(\omega')-1](V/c)\omega_0. \tag{15}$$

The above shift thus has the same sign as $V$ if $n(\omega') > 1$, otherwise it is an inverse Doppler shift, although it is hard to associate any practical significance with the Doppler shift inside a moving slab as seen by a stationary observer! Finally, we examine the $E$- and $H$-fields inside the slab, as seen by the stationary observer in the $xyz$ frame. Using the Lorentz transformation rules $E_x = \gamma(E_x' + VB_y')$ and $H_y = \gamma(H_y' + VD_x')$, we find

$$E_0^{(t)} = \left[\frac{1+n(\omega')(V/c)}{1+(V/c)}\right][1+r(\omega')]E_0, \tag{16a}$$

$$H_0^{(t)} = \left[\frac{1+n(\omega')(V/c)}{1+(V/c)}\right][1-r(\omega')]H_0. \tag{16b}$$

When $n(\omega') < 1$, the material must be dispersive, otherwise the light pulse will propagate faster than $c$, which is against special relativity. However, in the case of $n(\omega') > 1$, it is possible to have a dispersionless medium, in which case the analysis of the pulse energy is fairly simple. Equations (16a) and (16b) then tell us that the fraction of incident photons that enters the slab is $1 - r^2(\omega')$, and that the energy content of the pulse inside the slab is equal to this number of photons times $\hbar\omega^{(t)}$. Note that this energy is no longer balanced by the incident minus reflected energy seen in the $xyz$ frame. The difference must be made up by the energy transferred to or from the (moving) slab.

**5. Negative-index media used in computer simulations**. The sections that follow contain the results of our numerical simulations of a moving mirror within (or immediately behind) a transparent NIM. To avoid multiple reflections, the entrance facet of the negative-index slab is coated with a quarter-wavelength-thick layer of a material having $\mu = -\sqrt{|\mu_{\text{slab}}|}$ and $\varepsilon = -\sqrt{|\varepsilon_{\text{slab}}|}$. Such a layer acts as a perfect antireflection (AR) coating at normal incidence at the chosen wavelength. The present section describes the model used for the NIM media, as well as certain technical aspects of our Finite Difference Time Domain (FDTD) simulations. The results of these simulations pertaining to the Doppler shift observed upon reflection from a moving object immersed in a transparent NIM will be discussed in Sec. 6. A related set of simulations involving a reflector detached from the NIM environment in which it is submerged will be the subject of Sec. 7.

For the negative-index medium that surrounds a moving object, we use the following Drude models for the frequency dependence of the permeability $\mu(\omega)$ and permittivity $\varepsilon(\omega)$:

$$\mu(\omega) = \mu_\infty - \frac{\omega_m^2}{\omega^2 - 2i\omega\delta_m}, \tag{17a}$$

$$\varepsilon(\omega) = \varepsilon_\infty - \frac{\omega_p^2}{\omega^2 - 2i\omega\delta}. \tag{17b}$$

In the reported simulations, $\mu_\infty = 1.0$, $\delta_m = 7.89889434 \times 10^{10}$ rad/s and $\omega_m = 4.20834989 \times 10^{15}$ rad/s, which yields a value of $\mu = -1.0$ at the vacuum wavelength of $\lambda_o = 633$ nm. For a NIM having $n = -1.0$ at $\lambda_o = 633$ nm, we set $\varepsilon_\infty = \mu_\infty$, $\delta = \delta_m$ and $\omega_p = \omega_m$, to result in $\mu = \varepsilon$ and an impedance $\sqrt{\mu/\varepsilon}$ of unity at all frequencies. For a NIM having $n = -1.5$ at $\lambda_o = 633$ nm, we multiplied the above $\varepsilon(\omega)$ by $1.5^2$, to get $\varepsilon_\infty = 2.25$, $\delta = 7.89889434 \times 10^{10}$ rad/s and $\omega_p = 6.31252483 \times 10^{15}$ rad/s. For these two cases, the real and imaginary parts of the complex refractive index $n + i\kappa = \sqrt{\mu\varepsilon}$ are plotted as functions of the vacuum wavelength $\lambda_o$ in Fig. 4. The imaginary part $\kappa$ is seen to be of the order of $10^{-4}$; our simulation results do not change appreciably even when we raise $\kappa$ to $\sim 10^{-3}$, so for the range of parameters used in our simulations, these media are essentially transparent.

In the case of the NIM having $n = -1.0$, no anti-reflection coating is necessary since the medium is automatically impedance-matched to free space at all frequencies. As for the AR layer applied to the entrance facet of the NIM having $n = -1.5$, we took $\varepsilon(\omega)$ used in the case of $n = -1.0$ and multiplied it by 1.5, which yielded $\varepsilon_\infty = 1.5$, $\delta = 7.89889434 \times 10^{10}$ rad/s and $\omega_p = 5.15415494 \times 10^{15}$ rad/s. Since we use the same $\mu(\omega)$ in all cases and only scale $\varepsilon(\omega)$ by a constant factor relative to $\mu(\omega)$, the admittances of the various NIMs used in our simulations are frequency independent. Moreover, in each case the admittance $\sqrt{\varepsilon/\mu}$ is equal to the absolute value of the refractive index $\sqrt{\mu\varepsilon}$ at $\lambda_o = 633$ nm.

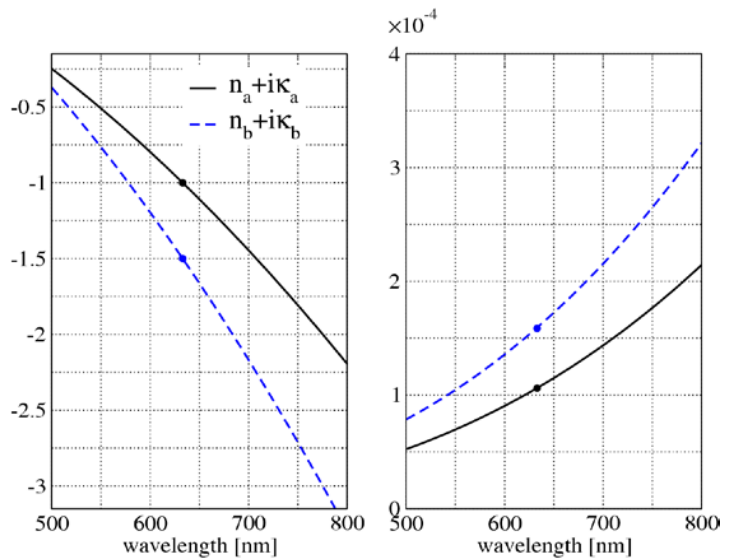

**Fig. 4**. Plots of the real part $n$ and imaginary part $\kappa$ of the complex refractive index versus the vacuum wavelength $\lambda_o$ for two negative-index media used in the simulations. The frequency dependences of the permeability $\mu(\omega)$ and permittivity $\varepsilon(\omega)$ are given in Eqs. (17). For the $a$ and $b$ media depicted here, the parameters $\mu_\infty$, $\delta_m$ and $\omega_m$ that specify $\mu(\omega)$ are the same. The parameters are chosen to yield $n_a = -1.0$ and $n_b = -1.5$ at $\lambda_o = 633$ nm. The imaginary parts $\kappa_a$ and $\kappa_b$, being of the order of $10^{-4}$, are negligible for all practical purposes. For the first medium, $\varepsilon_a(\omega) = \mu_a(\omega)$ at all frequencies, which results in an admittance $\sqrt{\varepsilon_a(\omega)/\mu_a(\omega)}$ of unity as well. For the second medium $\varepsilon_b(\omega) = 2.25\varepsilon_a(\omega)$, so its refractive index at $\lambda_o = 633$ nm is $n_b = -1.5$, while its admittance everywhere is 1.5.

In the simulations of the following sections, the submerged reflector is a positive-index dielectric slab of refractive index $n_1 = 2.0$ contiguous with the host medium of refractive index $n_o$. The rear facet of the reflector makes contact with the perfectly-matched boundary layer of the FDTD mesh, which acts as a perfect absorber, simulating a non-reflecting, open boundary condition. To simulate the motion of the reflector along the positive $z$-axis, we continually reduced the refractive index $n_1$ of the mirror's front facet (a single sheet of pixels within the FDTD mesh) until it became equal to the refractive index $n_o$ of the host dielectric. At this point the second layer of the mirror becomes its new front facet, whose refractive index must then be gradually reduced toward that of the host dielectric. The process continues at a fixed rate, with successive layers of the mirror transformed into the material that forms the incidence medium. Meanwhile the incoming light pulse continues to strike the shifting surface of the mirror, reflecting off the host/mirror interface, and returning to the incidence medium, where it now propagates in the negative $z$ direction.

We mention in passing that numerical inaccuracies arise if the mirror velocity is too small compared to the speed $c$ of light in vacuum. However, for mirror velocities greater than about $10^{-4}c$, FDTD simulations yield highly accurate values for the Doppler shift. These simulations evolve electromagnetic fields on a discrete grid in space and time. A small mirror velocity leads to a small difference between the incident and reflected wave frequencies. In order to resolve (numerically) this small difference in frequency caused by the Doppler shift, a long simulation time is required when monitoring the fields in time. (Recall that resolution in frequency domain is inversely proportional to the length of the sampling interval in time domain.) Spatial resolution must be increased accordingly as well, in order to keep numerical errors caused by spatial discretization from dominating the small frequency shifts that are being measured. Fortunately, since our theoretical results are known to be accurate for small mirror velocities, we need the FDTD simulations to ascertain only the upper limit of velocities at which the theoretical results can be relied upon. As it turns out, the simulations confirm the validity of the theory even at velocities approaching $10^{-2}c$, which is far greater than anything that we need for the application of our Doppler shift formulas to radiation pressure problems.

Note that, in computer simulations, it is possible to successively replace layers of the mirror material with those of the host medium (or vice versa), thus simulating a smooth and uniform motion of the mirror without complications arising, for example, from laminar or turbulent flow, from induced density gradients, or from elastic wave generation and propagation within the host dielectric. Such problems, however, inevitably arise in an experimental setting, which would involve motion of the host liquid, interactions between the light pulse and the moving liquid (including exchanges of energy and momentum), and inaccuracies in the applicable Doppler shift formula caused by departures from uniformity of the liquid as well as its motion. To avoid such complications, we have chosen to analyze in Sec. 7 the case of a moving mirror detached from a solid, rigid dielectric block.

**6. Submerged mirror in a negative-index medium**. With reference to Fig. 5, let the incident $E$-field amplitude referred to the origin of the coordinate system be $E_o\hat{x}\exp[i(k_0z-\omega_0t)]$, where $\omega_0$ is the optical frequency of the source and $k_0=n_0\omega_0/c$ is the wave-number inside the slab of refractive index $n_0$. If the interface between this dielectric slab and the reflector happens to be at $z=z_o$, the reflected $E$-field amplitude will be $rE_o\hat{x}\exp\{i[k_0(2z_o-z)-\omega_0t]\}$, where $r=(\eta_0-\eta_1)/(\eta_0+\eta_1)$ is the Fresnel reflection coefficient at the interface between media having admittances $\eta_0$ and $\eta_1$. Suppose the reflector now moves at a constant velocity $V$ along the $z$-axis, while remaining in contact with the dielectric host at all times. We write $z_o=Vt$ and assume that the reflection coefficient $r$ is independent of the velocity $V$. The reflected $E$-field amplitude thus becomes $rE_o\hat{x}\exp\{-i[k_0z+\omega_0(1-2n_0V/c)t]\}$, which clearly has a Doppler shift of $\Delta\omega/\omega_0=-2n_0V/c$. For $V>0$, this will be a red-shift for a positive-index host medium (i.e., $n_0>0$), and a blue-shift for a negative-index host.

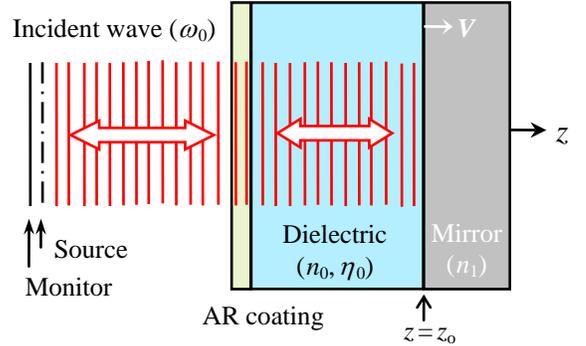

**Fig. 5**. Monochromatic plane-wave is reflected from a flat mirror submerged within a host medium of refractive index $n_0=-1.5$. The front facet of the NIM is antireflection coated, the host and the mirror are in contact at all times, and the mirror moves at a constant velocity $V$ along the $z$-axis. In our simulations, the mirror is a non-magnetic dielectric slab having $\mu_1(\omega)=1.0$ and $\varepsilon_1(\omega)=4.0$, i.e., refractive index $n_1=2.0$ and admittance $\eta_1=2.0$, whose rear facet is contiguous with the perfectly-matched boundary layer of the FDTD mesh.

The above theoretical argument is inexact for at least three reasons: (i) it is non-relativistic; (ii) the assumption that the reflection coefficient $r$ is independent of the velocity $V$ is unjustifiable; and (ii) the wave number of the reflected beam continues to be $k_0=n_0\omega_0/c$, while its frequency has shifted from $\omega_0$ to $\omega_0(1-2n_0V/c)$. It is thus necessary to verify the accuracy of the Doppler formula $\Delta\omega/\omega_0=-2n_0V/c$ using the essentially exact solution of Maxwell's equations provided by the FDTD numerical simulations. Figure 6 shows the simulated Doppler shift $\Delta\omega=\omega_r-\omega_0$ versus the time $t$ for a NIM host of refractive index $n_0+i\kappa_0=-1.5+1.6\times10^{-4}i$ under two different mirror velocities: $V_1=3\times10^{-3}c$ and $V_2=10^{-2}c$. In the latter case, the phase of the reflected beam was monitored once in the air (red curve) and a second time inside the NIM (green curve). When the monitor is placed inside the NIM, it is necessary to subtract the incident field from the total field at the location of the monitor. For this purpose, we did an additional simulation in which the $n_1 = 2.0$ reflector was removed while the $n_0 = -1.5$ slab was AR coated on *both* entrance and exit facets. We used the difference in the time dependences of the *E*-fields obtained with and without the reflector to compute the phase of the reflected field. The slope of this phase as a function of time then yielded the frequency $\omega_r$ of the reflected beam.

In all cases, we obtained for the reflected light a linearly increasing phase as a function of time. The derivative of the phase with respect to time, i.e., $\omega_r$, has some noise that oscillates around a constant average value. A fit of a straight line to the simulated data confirms that the average is indeed constant to good precision, and the value of $\omega_r$ computed from that constant is within 0.1% of the theoretical value computed from the formula $\omega_r - \omega_0 = -2n_0\omega_0 V/c$. While there is noise in the data that seems to increase with $V/c$, the result is clear: the sign of the Doppler shift is reversed in the case of the mirror immersed in a NIM, independently of whether the reflected signal is monitored inside the NIM or outside.

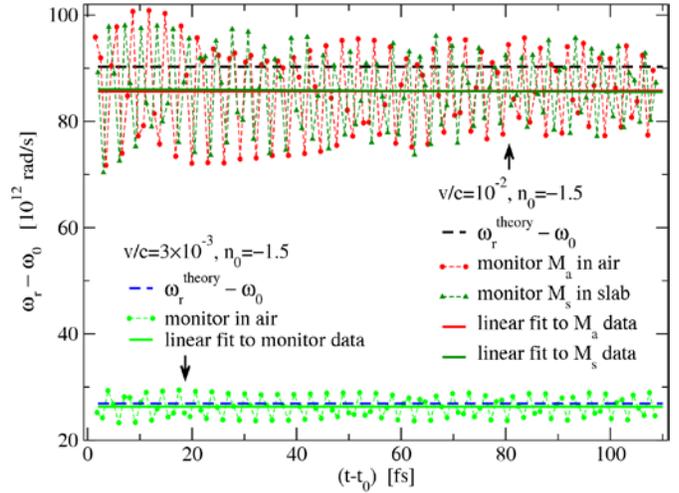

**Fig. 6**. Simulated values of the Doppler shift upon reflection from a moving dielectric of refractive index $n_1 = 2.0$ immersed in a NIM of refractive index $n_0 + i\kappa_0 = -1.5 + 0.00016i$ at two different velocities: $V = 10^{-2} c$ (top) and $V = 3 \times 10^{-3} c$ (bottom). The entrance facet of the host dielectric is antireflection coated. The phase monitor of the reflected light may be placed in the air (i.e., the incidence medium, where the source is located) or inside the NIM, without affecting the end result.

Note, once again, that in these FDTD simulations, as the reflector in Fig. 5 moves to the right at a constant velocity, the refractive index of the medium at the interface between host and reflector is continually changed from $n_1$ to $n_0 + i\kappa_0$. In other words, there is no fluid flow or turbulence or stretching of the host medium, etc., as there would be in a real physical situation. The FDTD method thus enables one to simulate an ideal motion of the reflector inside a dielectric host, without causing any of the above disturbances which would surely contaminate the results of any realistic experiment.

**7. The case of a mirror detached from the host dielectric**. In order to relate the Doppler shift of the light reflected from (and also transmitted into) a submerged partial reflector to the radiation pressure on the reflector, it is best to assume that the host is a massive, solid dielectric, detached from the mirror by a small air-gap. Such a system is depicted in Fig. 7, where a dielectric slab of refractive index $n_0$ and admittance $\eta_0$ is separated by a gap $d$ from a moving partial reflector. As before, the reflector is assumed to be a dispersionless, non-magnetic dielectric whose (positive) refractive index $n_1$ is equal to its admittance $\eta_1$, and whose rear facet is contiguous with the perfectly-matched absorbing boundary layer of the FDTD mesh. The front facet of the host medium is anti-reflection coated to avoid multiple internal reflections.

At the rear facet of the dielectric slab in Fig. 7 (i.e., at $z = 0$), one can readily derive an analytic expression for the Fresnel reflection coefficient $r$ as a function of $\eta_0$, $n_1$, $d$ and $\omega_0$. When the mirror moves at the constant velocity $V$, the gap-width $d$ becomes a linear function of time, namely, $d = d_o + Vt$, in which case the phase of the reflection coefficient $r$ can be differentiated with respect to time to yield an analytic expression for the Doppler shift $\Delta\omega = \omega_0 - \omega_r$ of the reflected beam. A similar procedure may be followed for the light transmitted into the mirror. Since the details of these calculations are given in [5] we will not repeat them here; suffice it to say that, at sufficiently small velocities, the effective Doppler shift – an average over reflected and transmitted photons – turns out to be $\Delta\omega = \alpha\omega_0 V/c$, where the coefficient $\alpha$ is a function of $\eta_0$, $n_1$, $\omega_0$ and the gap-width $d$ at the instant of observation $t$.

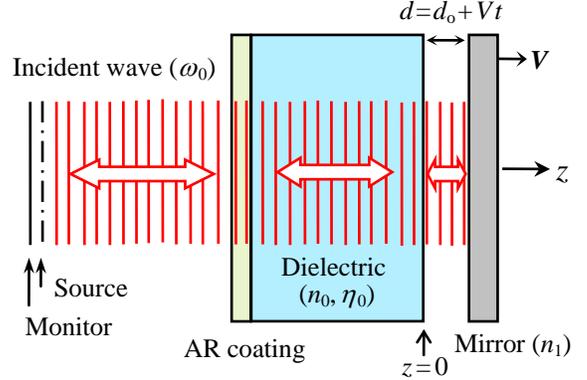

**Fig. 7** A monochromatic plane-wave is reflected from a flat mirror located behind a transparent slab of refractive index $n_0$ and admittance $\eta_0$. The front facet of the slab is anti-reflection coated, the initial air-gap between the mirror and the slab is $d_o$, and the mirror moves at a constant velocity $V$ away from the slab. In our simulations, the mirror is a non-magnetic dielectric slab of refractive index $n_1 = 2.0$, whose rear facet is contiguous with the perfectly-matched boundary layer of the FDTD mesh. Since the source in FDTD simulations radiates only in the positive $z$-direction, placing the monitor before the source ensures the capture of the reflected light without interference from the incident light.

Now, suppose a photon of energy $\hbar\omega_0$ enters the system of Fig. 7 where the reflector is initially at rest. The kinetic energy gained by the mirror, $\tfrac{1}{2}MV^2$, when equated to the average energy lost by the photon, $\tfrac{1}{2}\hbar\alpha\omega_0 V/c$, yields the acquired momentum of the mirror as $MV = \alpha\hbar\omega_0/c$. This acquired momentum clearly depends, through $\alpha$, on the admittance $\eta_0$ of the host medium, which is always a positive number, unlike the refractive index $n_0$, whose sign could be positive or negative. In the foregoing analysis, the gap-width $d$ may be made as small as desired, even zero, provided that the mirror is not allowed to stick to the dielectric host. Any difference between the change in the photon momentum (before and after passage through the system) and the momentum transferred to the mirror (i.e., $\alpha\hbar\omega_0/c$) is picked up by the host dielectric. Since the host is assumed to be rigid and massive, its acquired kinetic energy, compared to the kinetic energy gained by the mirror, will be negligible; this, of course, is an important requirement for the validity of the preceding analysis.

Our detailed analysis in [5] suggested that the *total* momentum per photon inside a transparent medium of admittance $\eta_0$ is $\tfrac{1}{2}(\eta_0 + \eta_0^{-1})(\hbar\omega_0/c)$. This momentum consists of an *electromagnetic* part, $\hbar\omega_0/(n_g c)$, and a *mechanical* part, which add up to yield the total momentum. Depending on the optical parameters of the system, the momentum (per incident photon) transferred to the reflector could be anywhere between $2\eta_0\hbar\omega_0/c$ and $2\hbar\omega_0/(\eta_0 c)$. Once again, the phase refractive index $n_0$, whether positive or negative, does not appear in these expressions; rather, it is the admittance $\eta_0$ and the impedance $\eta_0^{-1}$ that play decisive roles.

In the remainder of this section we present the results of FDTD computer simulations which confirm that (i) the Doppler shifts observed for positive- and negative-index dielectrics in the system of Fig. 7 are in fact similar to each other, and (ii) the magnitude of the shift depends on $\eta_0$, $n_1$, $d$, $\omega_0$ and $V$, but not on $n_0$. As before, the reflector used in the simulations was a non-magnetic dielectric slab having $n_1 = 2.0$, whose rear facet was rendered invisible by seamlessly merging into a perfectly-matched boundary layer.

In Fig. 8 we show the computed Doppler shift $\Delta\omega = \omega_r - \omega_0$ versus delayed time, $t - t_o$, with $\lambda_0 = 633$ nm, $n_1 = \eta_1 = 2.0$, $d_o = 100$ nm, and three different values for $(n_0, \eta_0)$. The delay $t_o$ is needed to allow the light pulse, which leaves the source at $t = 0$, to stabilize after traveling to the mirror, bouncing back, and returning to the monitor. The mirror velocity is $V = 3 \times 10^{-3} c$ in Fig. 8(a) and $V = 10^{-2} c$ in Fig. 8(b). Three sets of curves appear in Fig. 8(a), each set containing the simulated Doppler shift as a function of delayed time $t - t_o$ (solid), and the corresponding theoretical estimate obtained from the phase of the Fresnel reflection coefficient $r$ calculated at $z = 0$ (dashed). The maximum difference between theoretical results and numerical simulations is ~1%. For the NIM having $(n_0, \eta_0) = (-1.0, 1.0)$, $\Delta\omega$ is seen to be time-independent. This is because the slab is impedance-matched to free space; there is no reflection from either facet of the slab and, therefore, no light to interfere with the light reflected from the mirror. The second set of curves in Fig. 8(a) represents a positive-index medium having $(n_0, \eta_0) = (1.5, 1.5)$. Here the Doppler shift is a function of time because the widening gap between the dielectric slab and the mirror affects the phase (and thereby the frequency) of the reflected beam. Finally, the third set represents a NIM having $(n_0, \eta_0) = (-1.5, 1.5)$. A longer delay ($t_o = 190$ fs) was needed in this case to achieve a time-harmonic signal, presumably because of the large group refractive index ($n_g \sim 4.5$) of the slab. Had we continued the simulations for a longer time, the two curves corresponding to $n_0 = 1.5$ and $n_0 = -1.5$ would have been seen to be identical, except for a lateral shift along the time axis, which is a manifestation of the different group velocities in the two media.

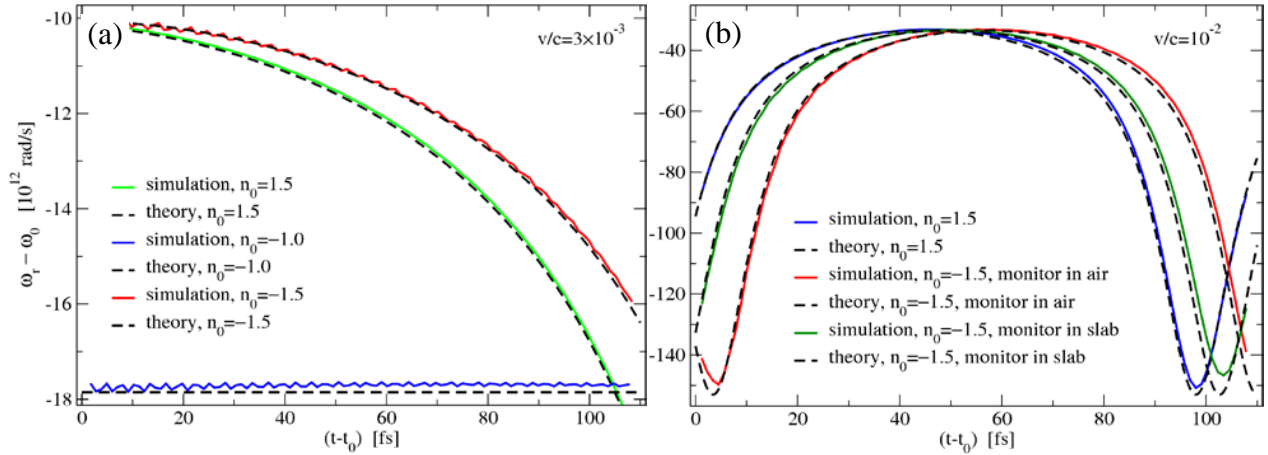

**Fig. 8**. Simulated values of the Doppler shift computed with a NIM slab of refractive index $n_0 + i\kappa_0 = -1.5 + 0.00016i$, with an anti-reflection layer at the entrance facet, and with a detached mirror of refractive index $n_1 = 2.0$ moving with constant velocity $V$ away from the slab. Also shown for comparison are simulated curves of $\Delta\omega$ versus time for a positive-index slab having $n_0 = 1.5$ and a NIM having $n_0 + i\kappa_0 = -1.0 + 0.00011i$. (a) $V = 3 \times 10^{-3} c$. (b) $V = 10^{-2} c$. Dashed black curves represent theoretical estimates, shifted along the horizontal axis in order to account for the time that it takes the light to propagate through the air and the slab, this time being dependent on the distance and the group refractive index of the media along the path from the mirror to the monitor. For the slab having $n_0 = -1.5$ and with the monitor placed in the air, the same transit time was used in (a) and (b) to shift the theoretical curves to their correct locations.

Figure 8(b) shows the simulated Doppler shifts (solid curves) as well as the corresponding theoretical estimates (dashed curves) for a mirror velocity of $V = 10^{-2} c$ for two transparent slabs, one having $(n_0, \eta_0) = (1.5, 1.5)$, the other having $(n_0, \eta_0) = (-1.5, 1.5)$. In the case of the NIM slab, the monitor was placed once in the air (red curve) and another time inside the slab (green curve); in both cases we used a longer delay, $t_o = 190$ fs, to reach the steady state. In the 105 fs interval during which the reflected beam was monitored, the mirror moved a distance of $\lambda_0/2$, after which the curves began to repeat themselves. The three curves of Fig. 8(b) are identical except for a lateral shift, which is due either to the difference in the group velocity between the two media, or to the different locations of the monitor. Note that the difference in the group refractive index $n_g$ between different media affects not only the required delay $t_o$ for the signal to reach steady-state at the location of the monitor, but also the gap-width $d$ at the time the monitored signal was interacting with the mirror. The horizontal shifts needed to bring into alignment the various curves of Fig. 8 are intended to synchronize the monitored signal with the corresponding gap width.

The transit time must be calculated based on the group velocity of the electromagnetic wave, not its phase velocity, which is negative (and therefore meaningless) in the case of NIM media. The group velocity $V_g$ is found from $V_g = c/n_g$, where $n_g = \mathrm{d}[\omega n_0(\omega)]/\mathrm{d}\omega|_{\omega=\omega_0}$. This formula is accurate for long pulses (i.e., pulses with a narrow spectrum). Therefore, to the extent that the slope of $\omega n_0(\omega)$ varies over the bandwidth of the pulse, there may be a slight difference between the calculated $V_g$ and the actual pulse velocity within the NIM slab. For the NIM having $n_0 = -1.5$, the real part of $n_g$ turns out to be ~4.5 at $\lambda_0 = 633$ nm. The theoretical curves (dashed black) in Figs. 8(a) and 8(b) are laterally shifted using transit times that are calculated based on the corresponding values of $V_g$.

**8. Summary and conclusions**. Individual photons of frequency $\omega$ carry energy in the amount of $\hbar\omega$ and, in vacuum, have a momentum along their propagation direction that is given by $\hbar\omega/c$. This knowledge, in conjunction with the laws of conservation of energy and momentum, may be used to derive expressions for the photon's Doppler shift upon interacting with moving (or movable) objects. Similarly, a knowledge of the Doppler shift, acquired by other means, may be used to arrive at expressions for the radiation pressure and photon momentum involved in light-matter interactions. In this paper we have derived several exact as well as approximate formulas for the Doppler shift of electromagnetic waves reflected from material media, in general, and from negative-index media, in particular. We have also verified, using FDTD numerical simulations representing exact solutions of Maxwell's equations, that our approximate Doppler-shift formulas are, in fact, highly accurate at non-relativistic velocities, even when these velocities approach a few percent of the speed $c$ of light in free space.

We argued that a recent claim of experimental observation of the "inverse" Doppler shift upon reflection from a NIM is incompatible with energy and momentum conservation laws. The proper experimental setting for such an observation would involve the motion of a reflector inside a stationary NIM environment. Such an experiment, while fairly easy to simulate with an FDTD code (as was demonstrated in Sec. 6), might face difficulties in practice as the moving mirror (or scatterer) should maintain contact with its NIM environment at all times, without inducing much motion in its surroundings, and without modifying the optical properties of its neighborhood in any significant way.

By introducing a small air gap between a mirror and its surrounding environment, one obtains an expression for the Doppler shift that no longer depends on the refractive index of the host medium; instead, the Doppler shift now becomes a function of the host's admittance, $\eta_0 = \sqrt{\varepsilon/\mu}$, a positive entity that cannot be used to distinguish positive- from negative-index media. The advantage of having an air gap is that it allows one to choose a rigid and massive host, which is needed to prevent any transfer of energy from the photons to the host medium. The Doppler shift then provides a direct measure of the kinetic energy acquired by the mirror in the process of reflection, which leads in a straightforward way to the momentum picked up by the mirror. Questions of radiation pressure and photon momentum can thus be answered rather trivially by first calculating the Doppler shift. We used this method in Sec.7 to argue that the photon momentum inside a transparent dielectric must be given by $\tfrac{1}{2}(\eta_0+\eta_0^{-1})(\hbar\omega_0/c)$, and that the momentum per photon picked up by a submerged reflector, depending on the system parameters, could be anywhere in the range from $2\eta_0\hbar\omega_0/c$ to $2\hbar\omega_0/(\eta_0 c)$.